\documentclass{www2010-submission}
\usepackage{lscape, graphicx}
\usepackage{rotating}
\usepackage{url}
\usepackage[center]{subfigure}

\begin{document}
\title{Memento: Time Travel for the Web}

%
%

\numberofauthors{3}
%


\author{
%
\alignauthor Herbert Van de Sompel\\
       \affaddr{Los Alamos National Laboratory,}
       \affaddr{NM, USA}\\
       \email{herbertv@lanl.gov}
\alignauthor Michael L. Nelson \\
       \affaddr{Old Dominion University,}
       \affaddr{Norfolk, VA, USA}\\
       \email{mln@cs.odu.edu}
\alignauthor Robert Sanderson\\
       \affaddr{Los Alamos National Laboratory,}
       \affaddr{NM, USA}\\
       \email{rsanderson@lanl.gov}
\and
\alignauthor Lyudmila L. Balakireva\\
       \affaddr{Los Alamos National Laboratory,}
       \affaddr{NM, USA}\\
       \email{ludab@lanl.gov}
\alignauthor Scott Ainsworth \\
       \affaddr{Old Dominion University,}
       \affaddr{Norfolk, VA, USA}\\
       \email{sainswor@cs.odu.edu}
\alignauthor Harihar Shankar \\
       \affaddr{Los Alamos National Laboratory,}
       \affaddr{NM, USA}\\
       \email{harihar@unm.edu}
}

\maketitle
\begin{abstract}

The Web is ephemeral. Many resources have representations that
change over time, and many of those representations are lost forever.
A lucky few manage to reappear as archived resources that carry
their own URIs.  For example, some content
management systems maintain version pages that
reflect a frozen prior state of their changing resources. Archives recurrently crawl the web to obtain the actual representation of resources, and subsequently make those available via
special-purpose archived resources.  In both cases, the archival
copies have URIs that are protocol-wise disconnected from the
URI of the resource of which they represent a prior state. Indeed, the
lack of temporal capabilities in the most common Web protocol, HTTP,
prevents getting to an archived resource on the basis of the URI
of its original. This turns accessing archived resources into a
significant discovery challenge for both human and software agents, which
typically involves following a multitude of links from the
original to the archival resource, or of searching archives for the
original URI. This paper proposes the protocol-based Memento solution
to address this problem, and describes a proof-of-concept experiment
that includes major servers of archival content, including Wikipedia
and the Internet Archive. The Memento solution is based on
existing HTTP capabilities applied in a novel way to add the temporal dimension. The result is a
framework in which archived resources can seamlessly be reached via
the URI of their original: protocol-based time travel for the Web.

\end{abstract}




\section{Introduction}
\label{sec:introduction}

``The web does not work,'' my eleven year old son complained. After checking power and network connection, I realized he meant something rather more subtle. The URI (http://stupidfunhouse.com) he had bookmarked the year before returned a page that didn't look like the original at all, and definitely was not fun. He had just discovered that the web has a terrible memory. 

Let us restate the obvious: the Web is the most pervasive information
environment in the history of humanity; hundreds of millions of
people\footnote{\scriptsize{http://www.internetworldstats.com/stats.htm}} access billions of resources\footnote{\scriptsize{http://googleblog.blogspot.com/2008/07/we-knew-web-was-big.html}} using a variety of wired or wireless devices.
The rapid growth of the Web was made possible by a suite of relatively
simple, yet powerful technologies including TCP/IP, URI, HTTP, and
HTML. The Web is also highly dynamic, with a significant percentage
of resources changing at different rates over time
\cite{1498837, fetterly:large-scale, koehler:web-page, 988674}.
Given the ubiquity of the Web, it is rather surprising to find how
poor its memory is regarding these continuous changes. Indeed, once
a resource has changed, accessing one of its prior versions
becomes a significant discovery challenge, no longer merely a matter of using
Web protocols to dereference its URI.

In essence, the time dimension is absent from the most common of Web
protocols, HTTP. This timelessness is even written into the W3C's
Architecture of the World Wide Web \cite{archWWW}, which reminds us that
dereferencing a URI yields a representation of the (current) state of the
resource identified by that URI, and highlights the impracticality of keeping
prior states accessible at their own distinct URIs:

\begin{quote}
Resource state may evolve over time. Requiring a URI owner to publish a new
URI for each change in resource state would lead to a significant number of
broken references. For robustness, Web architecture promotes independence
between an identifier and the state of the identified resource.
\end{quote}

Nevertheless, the Web does contain a meaningful amount of records of the past.
Sites based on Content Management Systems (CMS) such as Wikimedia,
the platform used by Wikipedia, keep the current version of a page
accessible at a generic URI, while older versions remain accessible
at version-specific URIs.  Special-purpose services that are concerned with
persistent referencing, such as WebCite, store a representation of the resource retrieved at the
time the service is invoked.  Also, inspired by the pioneering
work of the Internet Archive, there is an ever-growing international
Web Archiving \cite{brown:archiving-websites, masanes:web-archiving-book} activity that consists of recurrently sending out
crawlers to take snapshots of Web resources, storing those in
special-purpose distributed archives, and making them accessible
through tools such as the Wayback Machine\footnote{\scriptsize{http://www.archive.org/}}.
Transactional archives \cite{fitch2003web} store every materially
different representation of a web server's resources as they are
being delivered to clients. Currently, their use is primarily restricted
to applications that need to meet special legal requirements, such
as keeping an exact record of what has been delivered to users of
an ecommerce or government site, and they are therefore typically
not openly accessible.  Exploratory work is ongoing regarding the
establishment of a peer-to-peer web archive that receives its content
from browser caches, and that therefore can be considered a client-side
transactional archive \cite{1555455}. Also personal client-side transactional archives have been proposed \cite{cooper:infomonitor, chungwwa:webarchiving, 1622221} but their private purpose excludes accessing them on the Web. Search engine caches may also contain prior representations of resources, but they are
restricted to the most recent snapshot taken by a crawler.

Although this variety of archival solutions exists and their coverage
is growing, accessing last year's version of a resource remains
a significant challenge. In the case of Wiki-\\pedia, one has to resort
to its History tab and navigate the sometimes thousands of entries
there. The situation is similar for most other version-aware sites.
For news sites, one may find the answer by searching the site's
special purpose archive if one exists. And, as an option of last
resort unknown to many Web users, one can individually search the
many Web Archives, hoping to find a page that was archived at a
time close to the desired one. This situation is cumbersome for
users who, for example, want to revisit a bookmarked resource as it
existed at the time of bookmarking. Research has indicated that anywhere between 50\% and 80\% of page visits are revisits \cite{1518909, Obendorf:chi07, tauscher1997}. To an extent, this finding
emphasizes the need for end-user time travel on the Web.

The poor integration of archival content in regular Web navigation
is also a fundamental hindrance to applications that require finding,
analyzing, extracting, comparing, and otherwise leveraging historical
Web information.  Examples include Zoetrope, a tool that allows
interaction with and visualization of high-resolution temporal Web
data \cite{1449756}; DiffIE, a Web browser plug-in that emphasizes
Web content that changed since a previous visit \cite{1622221}; and time-oriented search that tracks the
frequency of words and phrases in resources over time
\cite{1557077}. These applications must 
build their own special-purpose archives in an ad-hoc manner in
order to achieve their goals.

In this paper, we present the Memento solution to allow temporal
access to the Web. Our solution is based on and is as simple
as the technologies that led to the rapid growth of the Web. It
focuses on seamless access to archival content (irrespective of its
location) as part of regular Web navigation for both human and software agents.
It does not deal with the aspect of creating, populating, and
maintaining archives, but rather leverages their existence. The
remainder of the paper is structured as follows: Section \ref{sec:conneg} briefly reviews transparent content negotiation for HTTP in order to allow a better understanding of Section \ref{sec:solution} which introduces the Memento solution for time travel on the Web; Section \ref{sec:experiment} describes an
experiment that provides a proof of concept for the solution; Section
\ref{sec:issues} discusses open issues; and Section \ref{sec:related}
provides an overview of related work; Section \ref{sec:conclusions}
holds our conclusion.

\section{Content Negotiation}
\label{sec:conneg}

Transparent Content Negotiation for HTTP \cite{rfc2295} (from here on abbreviated as \textit{conneg}) allows a client to select which representation it wants to retrieve from a transparently negotiable
resource; that is, a resource that has multiple representations (variants)
associated with it, each of which is available from a variant resource.
Currently deployed dimensions that are open to conneg are media type, language, compression, and character set. A client expresses preferences, possibly according to multiple dimensions, in special-purpose HTTP Accept headers. Preferences are qualified with ``quality'', or ``q'', values, that have a normalized value of 1.0 -- 0.0 (an argument without a q value is assumed to have q=1.0).  For example, by using the header ``Accept-Language: en, fr;q=0.7'' the client indicates that English is preferred and French is acceptable. Based on information in these headers, a server will either:

\begin{itemize}

\item Select an appropriate representation: There are two ways for a server to do so. One way is to provide a ``HTTP 200 OK'' response with a ``TCN: Choice'' header, and a Content-Location header that indicates the URI of the variant resource that delivered the representation. The other is to provide a ``HTTP 302 Found'' response with a ``TCN: Choice'' header, and a Location header that indicates the URI of where the client can access the variant resource.

\item Respond with a ``HTTP 406 Not Acceptable'' response if the server cannot meet the client's preferences as stated in the request.  The server then also returns a ``TCN: List'' header and a list of variant resources it possesses that are associated with the requested resource.  The client can then make an informed decision about variant selection.\footnote{\scriptsize{The client can also force a `` HTTP 300 Multiple Choices'' response \\by issuing a ``Negotiate: 1.0'' request header.  This rarely occurs in practice, but the response is functionally equivalent to a ``HTTP 406 Not Acceptable'' response.}}

\end{itemize}

RFC 2295 proposes a format for these lists, expressed as an Alternates response header that can be used in both the Choice and List scenarios. Web
servers do not necessarily support all the negotiation dimensions for all of
their resources, but do indicate the supported dimensions to clients (e.g.,
``Vary: negotiate, accept-language'' if the language dimension is supported).
Also, note that according to RFC 2295, variant resources do not themselves
support content negotiation\footnote{\scriptsize{Servers must return a ``HTTP 506 Variant Also Negotiates'' response if variant resources support conneg.}}.

As an example, presume a transparently negotiable resource
http://an.example.org/paper for which the following variant resources
are available: the paper in HTML and English (paper.html.en), in PDF and English
(paper.pdf.en), and in PDF and French (paper.pdf.fr).
Now presume a client wants to access the paper and has a preference for HTML and English. The interaction, in which the server makes a choice that fully honors the client's preferences, would then be (only headers relevant for conneg are shown):

\begin{scriptsize}
\begin{verbatim}
GET /paper HTTP/1.1
Host: an.example.org
Accept: text/html, application/pdf;q=0.8
Accept-Language: en-US, fr;q=0.7, de;q=0.5

HTTP/1.1 200 OK
TCN: choice
Vary: negotiate, accept, accept-language
Content-Location: /paper.html.en
Content-Type: text/html
Content-Language: en
Alternates: 
  {"paper.html.en" 1.0 {type text/html} {language en}},
  {"paper.pdf.en" 0.8 {type application/pdf} {language en}}, 
  {"paper.pdf.fr" 0.6 {type application/pdf} {language fr}}
\end{verbatim}
\end{scriptsize}

However, if the client prefers PDF over HTML and insists only on German language documents (French and English have q=0.0), the interaction in which the server cannot honor the request, and leaves the choice to the client would be:

\begin{scriptsize}
\begin{verbatim}
GET /paper HTTP/1.1
Host: an.example.org
Accept: application/pdf, text/html;q=0.8
Accept-Language: de, fr;q=0.0, en-US;q=0.0

HTTP/1.1 406 Not Acceptable
TCN: list
Vary: negotiate, accept, accept-language
Alternates: {"paper.pdf.fr" 0.8 {type application/pdf} 
  {language fr}}, {"paper.html.en" 0.5 {type text/html} 
  {language en}}, {"paper.pdf.en" 0.4 
  {type application/pdf} {language en}}
\end{verbatim}
\end{scriptsize}

\section{The Memento Solution}
\label{sec:solution}

In this section, we introduce the two core building blocks of the Memento
solution to allow temporal navigation of the Web: HTTP content negotiation
in the datetime dimension, and an API for archives of web resources that allows requesting an inventory of available archived resources associated with a resource with a given URI. 

\subsection{A Memento: An Archival Resource}

We introduce the term \textit{Memento} to refer to an archival record of
a resource.  More formally, a Memento for a resource URI-R (as it
existed) at time t$_i$ is a resource URI-M$_i$[URI-R@t$_i$] for which
the representation at any moment past its creation time t$_c$ is the
same as the representation that was available from URI-R at time t$_i$,
with t$_c$ $\geq$ t$_i$. Implicit in this definition is the notion that,
once created, a Memento always keeps the same representation.

In the remainder of this paper, the term \textit{original resource} is used to refer to a resource that itself is not a Memento of another resource, and URI-R is used to denote its URI. URI-M is used to denote the URI of a Memento.

\subsection{HTTP Datetime Content Negotiation}

We introduce the notion of content negotiation in the datetime dimension (from here on abbreviated as \textit{DT-conneg}), allowing a client to indicate that it is looking for past rather than current representations of a resource. This is achieved by using a special-purpose Accept header, experimentally named X-Accept-Datetime, which has datetimes (rather \\than  media type or similar) as its value: 

\begin{scriptsize}
\begin{verbatim}
X-Accept-Datetime: {Sun, 06 Nov 1994 08:49:37 GMT}
\end{verbatim}
\end{scriptsize}

Generally speaking,
DT-conneg works in very much the same way as existing conneg approaches: If a client wants to retrieve a Memento of the original resource URI-R, it
issues an HTTP GET at URI-R using the X-Accept-Datetime header to express
the datetimes of the archival record(s) of URI-R in which it is interested. The server
handling this HTTP GET request tries to honor it by delivering a
representation it chooses based on the client's datetime preference(s), and/or
by providing the client with a list of available variant resources, each of which is a Memento of URI-R.  Described in more detail below, two distinctions exist between DT-conneg and other conneg approaches:

\begin{itemize}

\item Cases exist in which the server hosting URI-R can not
itself honor the DT-conneg request, but instead redirects to
a server that can.

\item The list of available variant resources can be too extensive to be
expressed in an Alternates header. In this case, a combination of a sizeable Alternates header listing variants centered on the requested datetime(s), and an HTTP Link header pointing at an extensive list of variants is used.

\end{itemize}

Before deciding on the X-Accept-Datetime header, we investigated possible alternatives that could be used in HTTP interaction. We decided not to use the ``features'' extensibility mechanism introduced by RFC 2295 because it is geared at the fine-grained specification of variant options (e.g., paper size, color depth) and hence is not suitable for something with the primacy of datetime. Also, the ongoing Media Fragment work of the W3C \cite{w3c:fragment}
is not applicable because it proposes expressing a segment of a resource (e.g., a region of an image, a section of a video) as a URI fragment. It does not deal with the notion of a resource that has changing representations over time.

\subsection{A TimeGate: A Resource Capable of DT-conneg}

We introduce the term \textit{TimeGate} to refer to a transparently
negotiable resource that supports the datetime dimension. More
formally, a TimeGate for an original resource URI-R is a transparently
negotiable resource URI- G[URI-R] for which all variant resources
are Mementos URI-M$_i$[URI-R@t$_i$] of the resource URI-R. Since
multiple archives may host versions of URI-R, multiple TimeGates
may exist for any given resource, i.e. one per archive.

\subsection{Time Travel: Combining DT-conneg and TimeGates}
\label{sec:headers}

To further explain DT-conneg and TimeGates, two separate scenarios
are explored. The combination of these scenarios provides a solution
for temporal Web navigation that integrates operational web servers
and archives of all types.  To allow for a better understanding,
the description is restricted to conneg in the datetime dimension only.
Also, in order to keep examples simple, requests with multiple datetime
values and associated q-values are not used.
It should be noted, however, that both multi-dimensional conneg, and multiple datetime values are possible in the proposed framework, since it builds on the principles of RFC 2295 that provides these capabilities.  Furthermore, we assume that the server that hosts the original resource URI-R for which a client wants to retrieve Mementos, is able to detect the existence of an X-Accept-Datetime header.

Before describing the scenarios, let us provide some explanatory information about the HTTP headers that are involved:

\textit{Alternates}: RFC 2295 requires listing all variant resources.
However, since an extensive set of variant resources may exist in
case of DT-conneg, the Alternates listing is impractical. Therefore,
Alternates only lists a limited amount of variant resources, centered
on the datetime requested by the client.

\textit{Link}: To compensate for the incomplete list of
variant resources in Alternates, an HTTP Link header \cite{HTTPLink} provides a pointer to
a resource (the TimeBundle, see Section \ref{sec:API}) that supports retrieving a list of
all variant resources (Mementos), and their associated metadata.

\textit{X-Archive-Interval}: Indicates the entire datetime interval
for which the archival server has Mementos for URI-R.

\textit{X-Datetime-Validity}: Indicates the datetime interval during
which the provided representation was valid. Certain servers,
including CMS and transactional archives, can reliably provide this
information. Others, such as crawler-driven web archives cannot.

\subsubsection{Web servers with archival capabilities}

Some web servers handle aspects of resource archiving natively, by maintaining explicit information about the location and datetimes of archival records of their resources, stored internally or remotely. Many CMS, Version Control Systems, as well as the TTApache
system \cite{dyreson:managing} fall under this category. But also
servers that recurrently archive into a cloud store and keep track of the URIs of the remote archival records fit in.

When a client is looking for Mementos of an original resource URI-R hosted
by these servers, they can handle the requests internally since all
the information that is required -- URIs of Mementos and their
datetimes -- is available.  In this case, the set-up is as follows:

\begin{itemize}

\item URI-R itself becomes a transparently negotiable resource that
supports DT-conneg to provide access to all its available Mementos.
In essence, URI-R functions as its own TimeGate URI-G.  Note that typical
URI-Rs for these systems either provide access to the current version
of a resource, or to a list of all its versions (each with its own
URI-M), or to a combination of both.

\item All Mementos URI-M$_i$[URI-R@t$_i$] of URI-R become variant resources
for URI-R.

\end{itemize}

\begin{sidewaysfigure*}
\begin{center}
\includegraphics[scale=0.42]{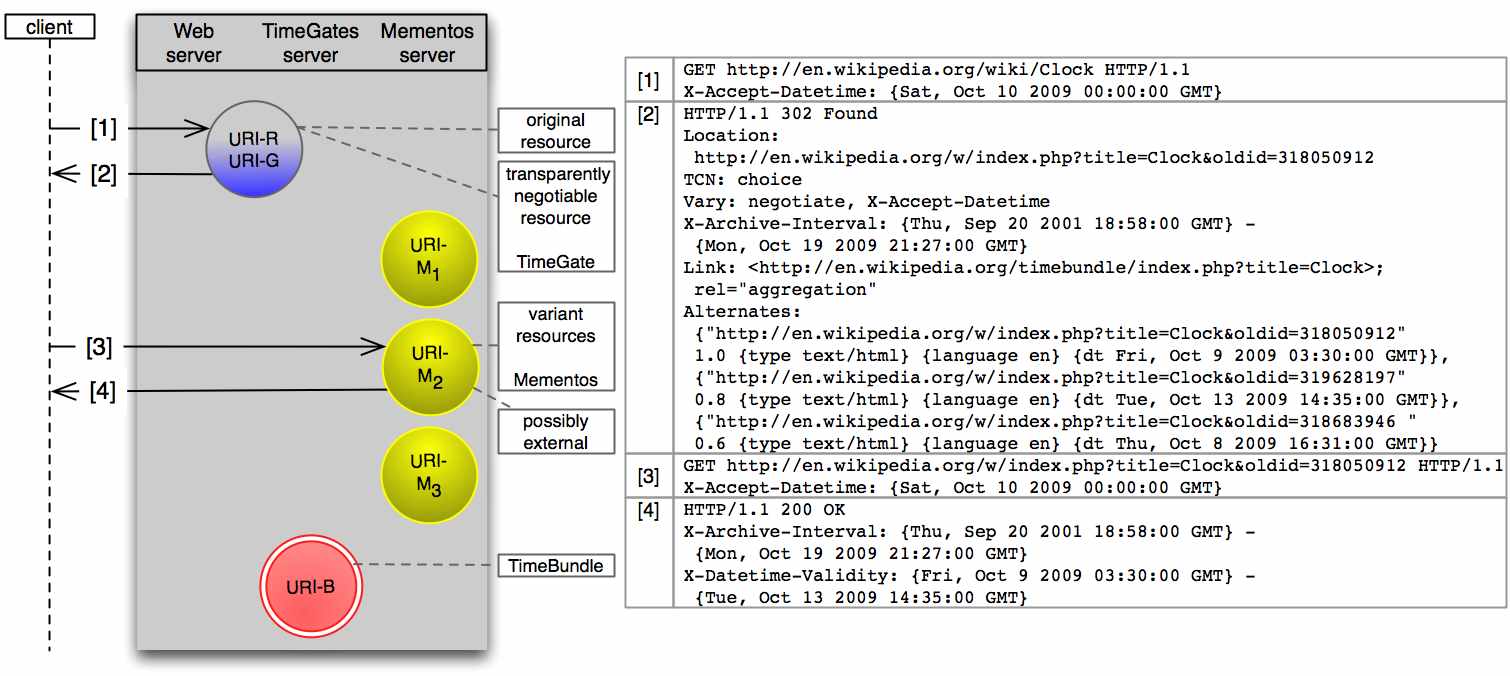}
\caption{DT-conneg for Web servers with archival capabilities: URI-R $=$ URI-G.}
\label{fig:scenario1}
\end{center}
\end{sidewaysfigure*}

Figure \ref{fig:scenario1} depicts a typical, successful, DT-conneg
transaction flow for this type of server, including the HTTP headers that are used. The transactional behavior for less trivial cases are also considered
in the Memento solution but space prevents us from
discussing them here.  Such cases include requesting Mementos for datetimes
that are out of the date-range for which the server has archival
records, requesting Mementos for URI-Rs that no longer exist, and the client providing a datetime which the server is unable to parse\footnote{\scriptsize{Details: {http://mementoweb.org/guide/http/local}}}.


\subsubsection{Web servers without archival capabilities}
\label{sec:nolocal}

Many other servers have no local archival capabilities \\whatsoever.
They host resources for which only a representation of the current
state can be retrieved, and are unaware of the details regarding
the existence of Mementos of their resources in other archival servers.  Naturally, such a server cannot redirect a client that
requests an archival record of one of its URI-Rs to an appropriate
Memento. However, these systems can still play a constructive role
by redirecting the client to a server that is equipped to handle
the request: an archive of web resources. In this case, the set-up is as follows:

\begin{itemize}

\item Upon detection of the X-Accept-Datetime header in the client's
request for URI-R, the server merely redirects (using ``HTTP 302 Found'') the client to an \\archival server. Note that this is not a 302 redirection that is part of a conneg transaction, as described in Section \ref{sec:conneg}. Rather it is a 302 redirection that results from detecting the X-Accept-Datetime header.

\item The redirection is to a TimeGate \\URI-G[URI-R] that the
archival server makes available for the original resource URI-R.

\item The archive's URI-G is a transparently negotiable resource
that supports DT-conneg to provide access to all the Mementos that
the archive has available for URI-R.

\item All Mementos URI-M$_i$[URI-R@t$_i$] that the archive has available
for URI-R become variant resources for its URI-G.

\end{itemize}

\begin{sidewaysfigure*}
\begin{center}
\includegraphics[scale=0.42]{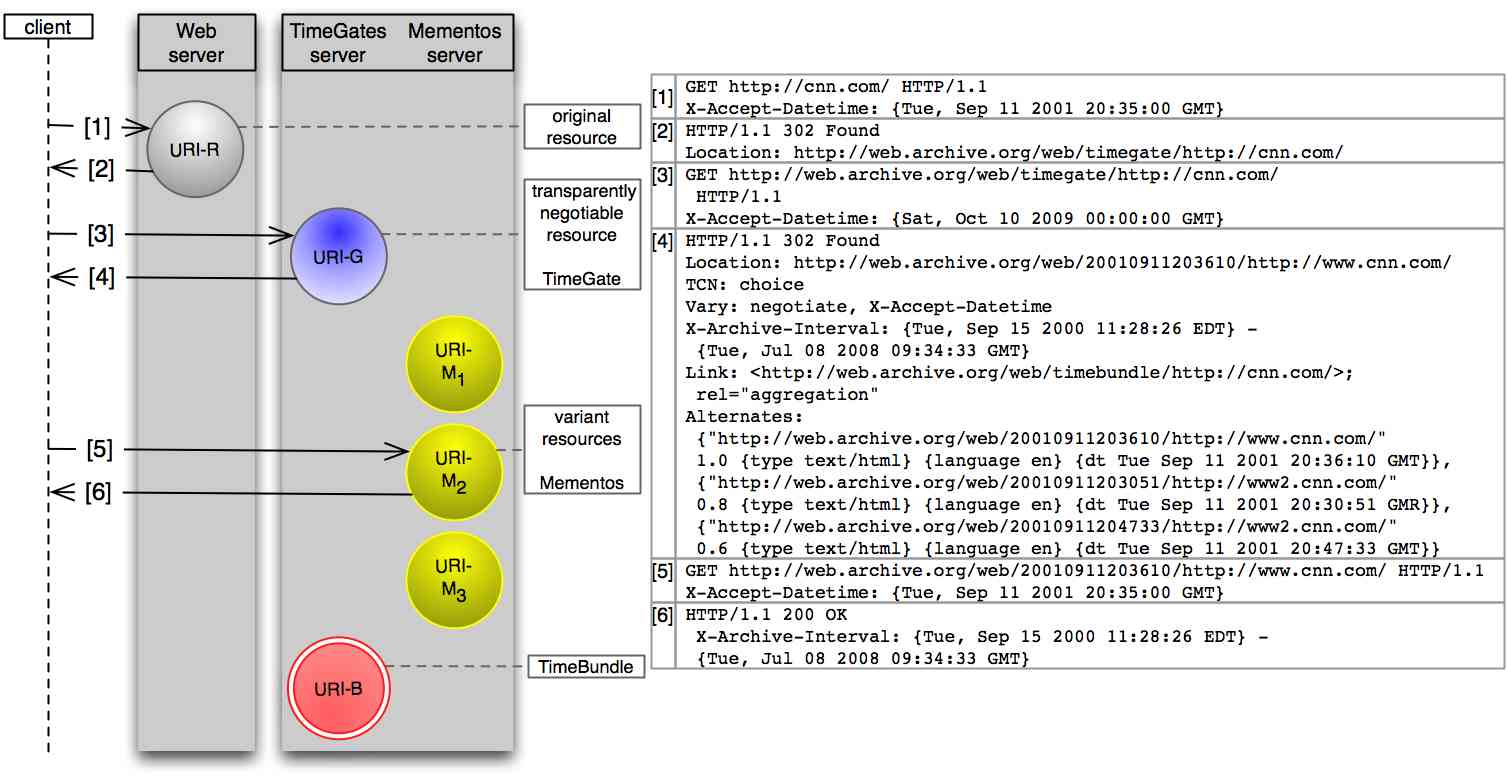}
\caption{DT-conneg for Web servers without archival capabilities: URI-R $\neq$ URI-G.}
\label{fig:scenario2}
\end{center}
\end{sidewaysfigure*}

Figure \ref{fig:scenario2} depicts a typical, successful, DT-conneg
transaction flow for this type of server, and includes the HTTP headers that are involved. Again, the transactional behavior for less trivial cases
is not covered here\footnote{\scriptsize{Details: http://mementoweb.org/guide/http/remote}}. In essence, the solution is the same as in the above case, with the exception that the TimeGates reside on an external archival server, not on the server that hosts the original resource URI-R. This distinction raises two important questions.

First, to which archive should a server redirect? In order to help the client, a
server should redirect to an archive that has the best archival coverage of its
resources. Servers that have an associated transactional archive should
redirect to it, servers that have explicit recurrent crawling agreements with
systems such as Archive-It\footnote{\scriptsize{http://www.archive-it.org/}} should point there, other servers may point at
their country-specific archive (such as the Finnish, Danish, Canadian, etc.
archives), and in many cases servers can point at the Internet Archive. Note
that scenarios may be envisioned in which the redirection is subject to
configuration, for example, redirection to different archives depending on
archival time-range, media type, etc. Then again, this problem of redirecting
to a specific archive could be addressed by uniformly pointing at an
aggregator service that holds crucial metadata (e.g., URI-R, URI-G, URI-M, t$_i$)
about Mementos available in a variety of archival servers, and that exposes
cross-archive TimeGates URI-G[URI-R]. In Section \ref{sec:API}, we introduce a
discovery API for archives that enables the creation of such a TimeGate
aggregator.

Second, how does the server know the URI-G of the \\TimeGate for its
own URI-R on an external archival server? This problem can be addressed
by introducing archive-\\specific or cross-archive conventions for
the syntax for URI-G of TimeGates as a function of URI-R. This would
simply formalize the status-quo as all major web archives that use the
Heritrix/Wayback solution already use such conventions.  For example, the
URI to retrieve a list of all archived versions of http://cnn.com/ is:\\

\begin{scriptsize}
\url{http://web.archive.org/web/*/http://cnn.com/}\\
\end{scriptsize}

Hence, the URI that could be used as a convention for the Internet
Archive's TimeGate for http://cnn.com/ would be:\\

\begin{scriptsize}
\url{http://web.archive.org/web/timegate/http://cnn.com/}\\
\end{scriptsize}

Such a convention seems achievable in the
context of the International Internet Preservation
Consortium\footnote{\scriptsize{http://www.netpreserve.org/}} that has
made archive interoperability one of its goals. However, when a TimeGate
aggregator service is introduced, URI-G syntax conventions for individual
archives are not crucial; only a convention for the aggregator's URI-G
syntax would be essential.


\subsection{Discovering Mementos: TimeBundles and TimeMaps}
\label{sec:API}

For discovery purposes, we introduce the notion of a resource hosted by an archival server, via which a full overview is available of all Mementos that the archive holds for an original resource URI-R; we name such a resource a \textit{TimeBundle}. More formally, a TimeBundle for a resource URI-R, is a resource URI-B[URI-R] that is an aggregation of: (a) all Mementos URI-M$_i$[URI-R@t$_i$] available from an archive, (b) the archive's TimeGate URI-G for URI-R, (c) the original resource URI-R itself.

Given the semantics of a TimeBundle, as an aggregation
of a set of resources, all of which share a temporal relationship with URI-R, we propose
to model it as an ORE Aggregation \cite{ldow2009:ore}. The ORE
specifications comply with the Linked Data conventions \cite{linkeddata}, and treat
an ORE Aggregation as a non-information resource \cite{w3c:HTTP14} described by an information resource that is accessible via
an HTTP 303 redirect from the URI of the ORE Aggregation. We name
the information resource that describes the TimeBundle a \textit{TimeMap};
it is a specialization of an ORE Resource Map. The TimeMap lists
the URIs of all resources that are aggregated in the TimeBundle,
as well as metadata that is available about them.  We have not
formally engaged in specifying which metadata to convey in TimeMaps,
but essentials such as archival datetime, media type, and language, as well as more specific information such as digest, number of observations, validity time-range \cite{1555455} must be considered\footnote{\scriptsize{An example RDF/XML TimeMap as used in our experiment is available at http://mementoweb.org/guide/api/map1}}.

TimeBundles made available by archives may be leveraged in real-time
client interaction, since their URI-B is expressed as the content of the
HTTP Link header (see the HTTP headers in Figures \ref{fig:scenario1} and \ref{fig:scenario2}).
And, when an archive makes its TimeBundles discoverable using common
approaches such as SiteMaps \cite{sitemaps}, Atom Feeds \cite{rfc:4287},
or OAI-PMH \cite{379449} they become a powerful mechanism for batch
harvesting of metadata that describes an archive's entire collection,
and that can be used for the creation of cross-archive services.

\subsection{A TimeGate Aggregator}

If various archives implement TimeBundles and associated TimeMaps,
and make them discoverable using the aforementioned techniques,
then information about Mementos hosted by different archives can be
harvested into an aggregator service.  For each original resource URI-R,
for which Mementos exist in the harvested archives, such an aggregator
then minimally holds the distinct URI-Ms of each of those Mementos in the
various archives, as well as their archival datetime, media type, language
etc. This information allows the aggregator to introduce TimeGates URI-G
for each of the URI-Rs for which the harvested archives have Mementos. The
variant resources for any specific TimeGate are the Mementos for URI-R
as they exist in the distributed archives.  Because the aggregator has
information on Mementos across archives, its time-granularity is finer
than that of any of the individual archives. This provides the aggregator
with a better range of possibilities when redirecting a client to a
Memento in response to a request for a specific datetime. In essence,
this aggregator behaves as the archival servers discussed in Section
\ref{sec:nolocal}, but it has a broader coverage both regarding URI-Rs
and Memento datetimes, and it does not store the Mementos itself.

\begin{figure*}
\begin{center}
\subfigure[Archive-It~~~~~~ \newline Thu, 08 Sep 2005 17:48:47 GMT]{\label{distributed-a}\includegraphics[scale=0.20]{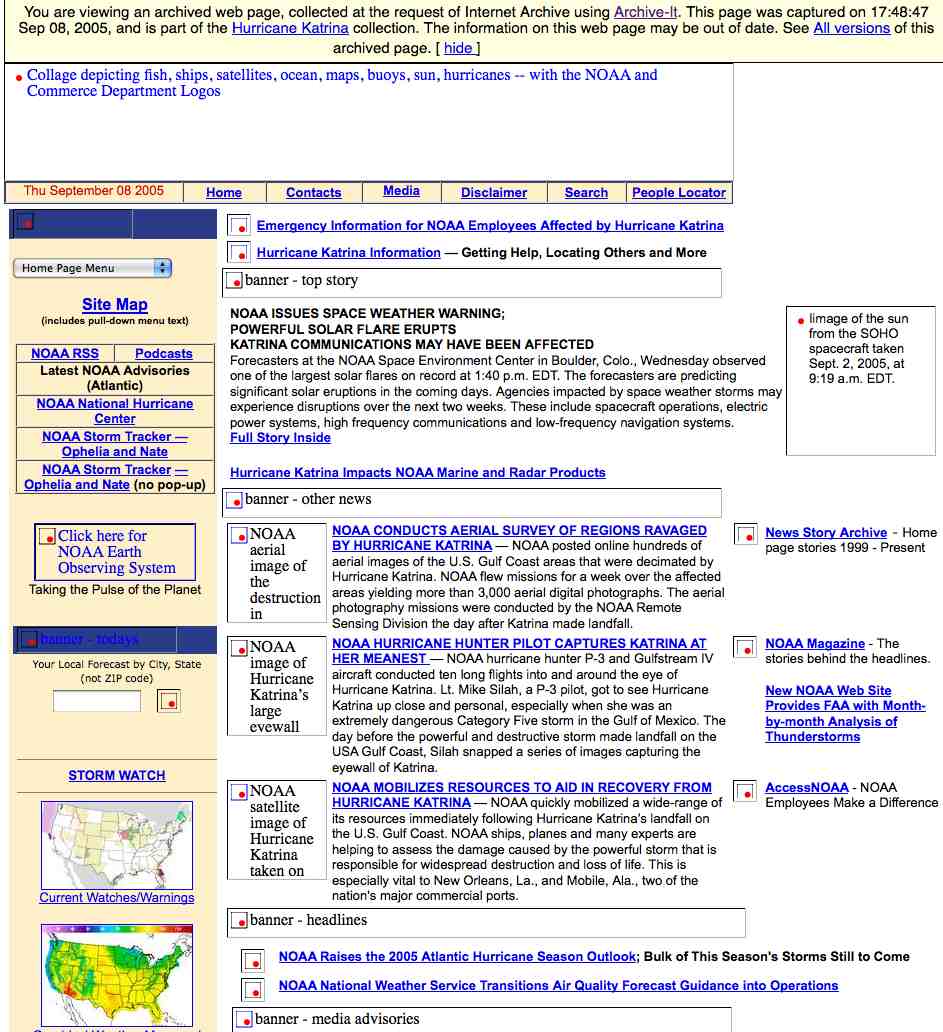}}
\subfigure[Internet Archive \newline Thu, 08 Sep 2005 21:07:05 GMT]{\label{distributed-b}\includegraphics[scale=0.20]{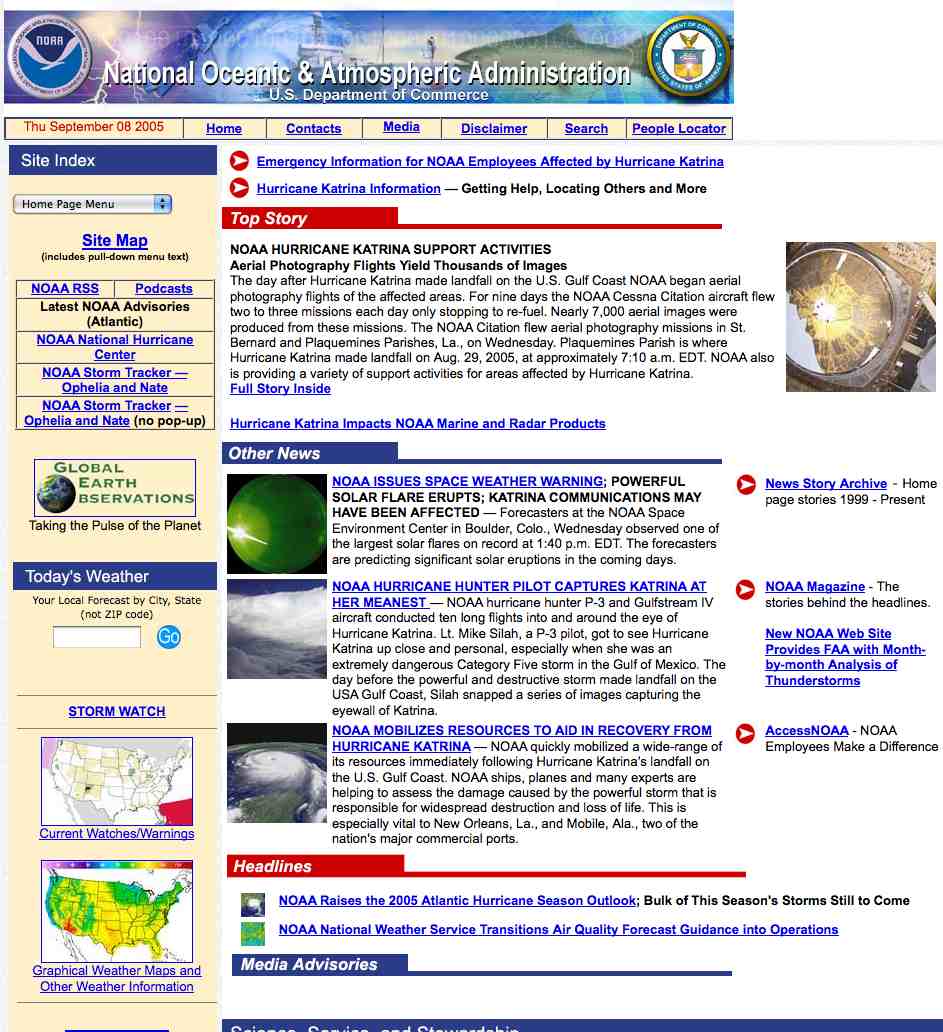}}
\subfigure[Internet Archive \newline Fri, 09 Sep 2005 01:58:48 GMT]{\label{distributed-c}\includegraphics[scale=0.20]{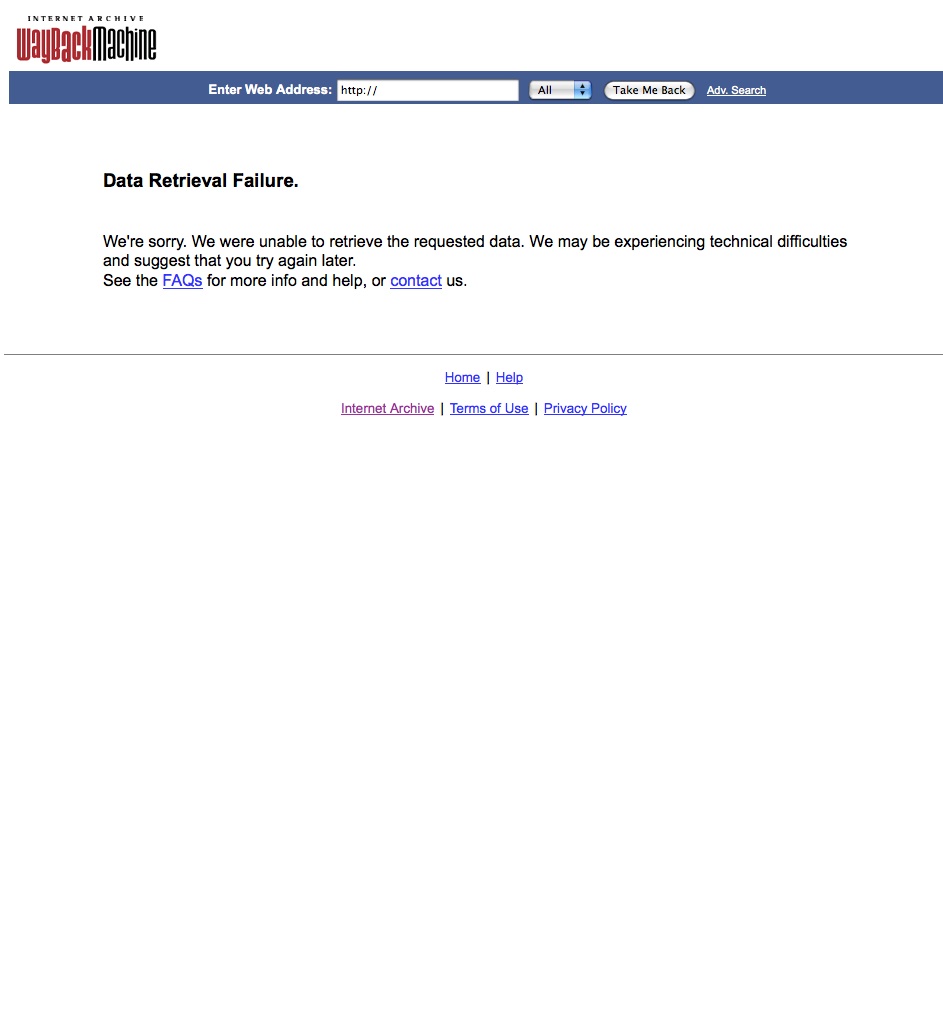}}
\subfigure[Archive-It~~~~~~ \newline Sat, 10 Sep 2005 08:11:47 GMT]{\label{distributed-d}\includegraphics[scale=0.20]{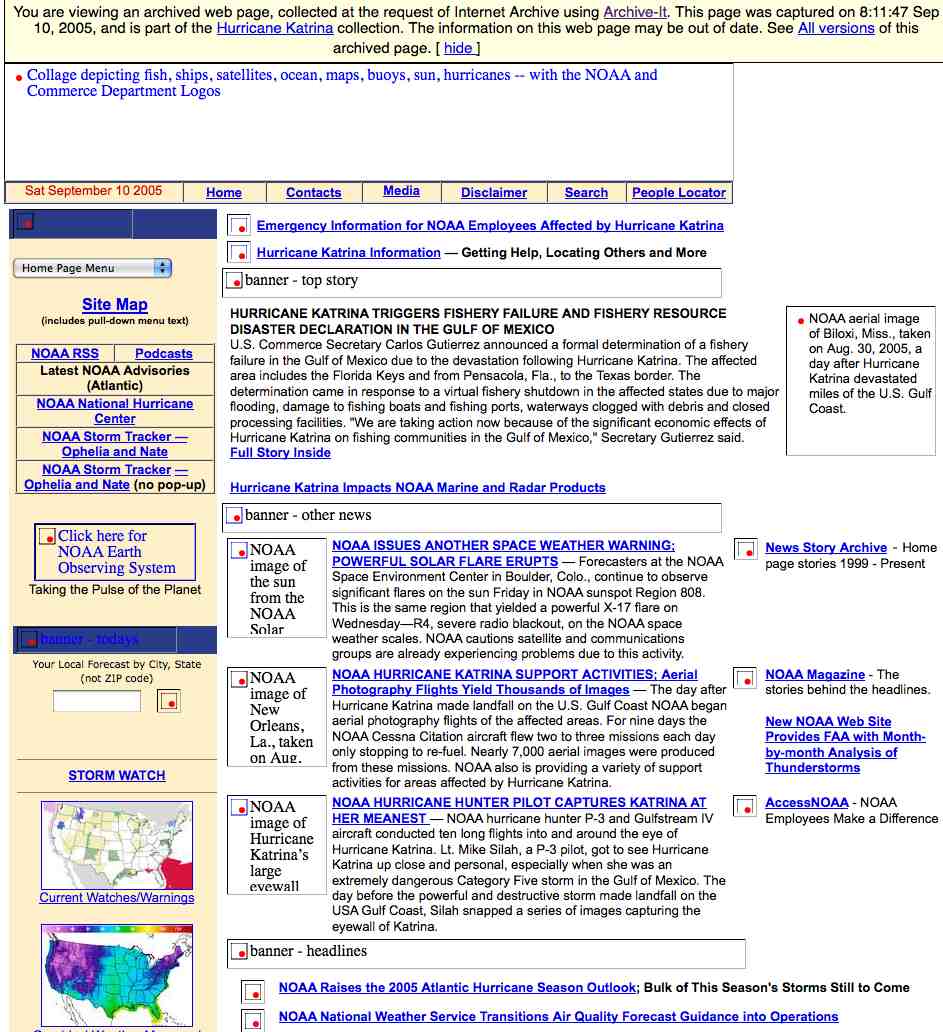}}
\caption{Distributed archive coverage of www.noaa.gov.  \ref{distributed-a} and \ref{distributed-d} come from Archive-It and \ref{distributed-b} and \ref{distributed-c} come from the Internet Archive.  Note that \ref{distributed-c} has either a transient or permament error.}
\label{fig:distributed}
\end{center}
\end{figure*}

Figure \ref{fig:distributed} illustrates the value such an aggregator
can bring to time travel. It shows various Mementos for the noaa.gov
home page as it was around the time of Hurricane Katrina. In order
to revive how the drama unfolded, inspecting Mementos held by different
archives is required. Indeed, both the content of the Mementos as well
as their archival server changes as time progresses. Note also that,
although the Internet Archive claims to have coverage for September
9 2005, the Memento is not really available (bottom left of Figure
\ref{fig:distributed}; it is not known if this is a permanent or
transient error); the next available Memento is for September 10
2005, and is available from Archive-It. In cases like this, an aggregator
could support navigation across archives and across time.


\section{Experiment}
\label{sec:experiment}

\begin{figure}[ht]
\begin{center}
\includegraphics[scale=0.25]{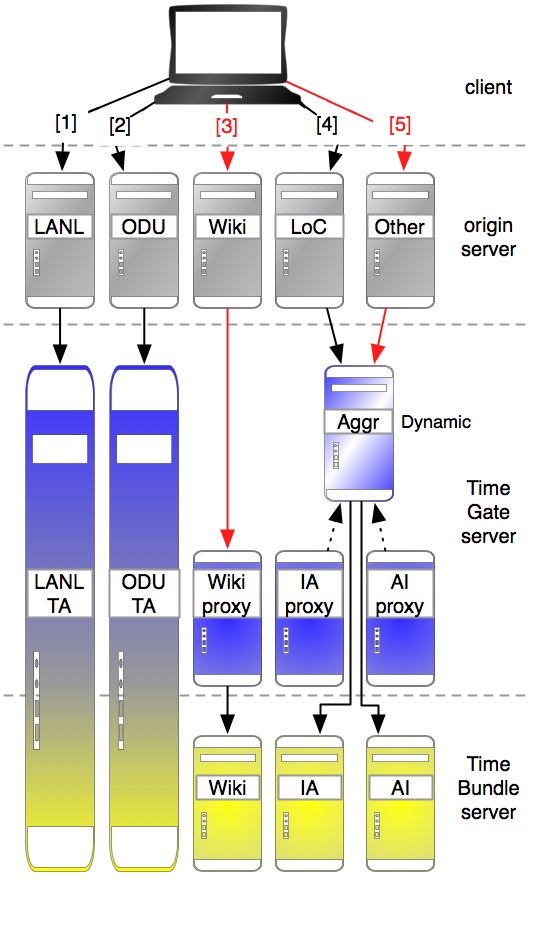}
\caption{The Memento Experiment environment.}
\label{fig:environment}
\end{center}
\end{figure}

We have performed an experiment to demonstrate the feasibility of the proposed DT-conneg framework involving a diverse array of components that jointly realize web time travel across various servers. The deployed environment is depicted in Figure \ref{fig:environment}. The arrows indicate the flow of HTTP interactions shown in Figures \ref{fig:scenario1} and \ref{fig:scenario2}, subject to the following considerations that are directly related to conducting a time travel experiment in a Web that is not (yet) DT-conneg enabled.

First, as it was not realistic to try and get active development involvement from existing archival servers within the timeframe the reported work took place, TimeGates and TimeBundles for several archives (CMS and web archives) were not implemented natively within those systems but rather by-proxy.  This means that they were exposed by servers under our control, which obtained the essential information from the archives using ad-hoc techniques such as screen scraping. While it may seem that this approach undermines the essence of the protocol-based DT-conneg framework, it actually is a strong illustration of its feasibility: if one can scrape the essential information from archives' pages, it is certainly available in their databases, and hence, native implementation should be more straightforward than by-proxy. Also, while we rely on a by-proxy approach for certain systems, demonstrations of the feasibility of native implementation are also available.

Second, existing web servers do not currently detect the X-Accept-Datetime header required for time travel, and hence cannot issue the essential ``HTTP 302 Found'' to a TimeGate (see Section \ref{sec:nolocal}). These servers will currently respond as usual, typically with an ``HTTP 200 OK'' or ``HTTP 404 Not Found''. In order to still be able to demonstrate the DT-conneg framework in the experiment, the remedy is to have the time travel client detect such responses that are unexpected from the time travel perspective, and take control by subsequently issuing the DT-conneg request directly to a TimeGate for URI-R exposed by an archival server (native or by-proxy).  In essence, in these cases, the client fulfills the redirecting role that the host of URI-R normally would in the DT-conneg framework. For servers outside of our control, there was no other option than to resort to this client approach; for servers under our control the redirect to a TimeGate was implemented natively. 

The following is a description of the components involved in the experiment:

\textit{Web servers}: We equipped domains under our own control with the
capability to honor DT-conneg requests by detecting the X-Accept-Datetime
header, and redirecting to TimeGates exposed by an appropriate archival
server.  This was trivially implemented using an Apache mod\_rewrite
\\rule\footnote{\scriptsize{See http://mementoweb.org/tools/apache}}
for servers we could configure:\\

\begin{scriptsize}
\noindent\url{http://lanlsource.lanl.gov/}\\
\url{http://odusource.cs.odu.edu/}\\
\url{http://digitalpreservation.gov/}\\
\end{scriptsize}

(LANL, ODU, and LoC, respectively in Figure \ref{fig:environment}). For
obvious reasons, we were not able to implement this for servers beyond
our control.

\textit{Archives}: Wikipedia is a prominent example of the class of servers with local archival capabilities. TimeGates (and TimeBundles) for it were implemented by-proxy (Wiki proxy in Figure \ref{fig:environment}). However, to demonstrate the possibility of native implementation, a plug-in was developed that adds X-Accept-Datetime and TimeGate capabilities to the Wikimedia platform on which Wikipedia is based\footnote{\scriptsize{Plug-in at {http://mementoweb.org/tools/wiki}}}. To cover for the class of servers that lack local archival capabilities, TimeGates (and TimeBundles) were implemented by-proxy for the Internet Archive (IA proxy in Figure \ref{fig:environment}), the Internet Archive's Archive-It (AI proxy in Figure \ref{fig:environment}), the Library of Congress' Archive-It, the Government of Canada Web Archive, and WebCite. In addition, we developed a transactional archive platform and deployed it at LANL and ODU (LANL TA and ODU TA in Figure \ref{fig:environment}, respectively). As the LANL and ODU web servers respond to client requests, the representations they serve are pushed into these archives, yielding a high-resolution archival record of their evolving resources. It is worth noting that the described selection covers a broad range of commonly deployed archival solutions: CMS, web-crawler based archives, on-user-demand archives, and transactional archives.

\textit{Aggregator}: Furthermore, a TimeGate aggregator (Aggr in Figure \ref{fig:environment}) was developed that collects archival metadata from the aforementioned web archives' TimeBundles (some by-proxy and some native), and can hence serve as a common target for redirection. This collecting is currently done dynamically: as a client requests a Memento for an original resource URI-R via the aggregator, the aggregator contacts associated TimeBundles in various archives, merges the returned TimeMap information, and only then redirects the client to an appropriate Memento. This experimental approach makes retrieving Mementos via the aggregator predictably slow.  

\textit{Clients}: We developed a FireFox plug-in that allows setting
the browser to time travel mode, and selecting a datetime for the
journey. From there onwards, the browser adds an X-Accept-Datetime
header, with the datetime value set by the user, to every HTTP
GET issued. If all targeted servers would implement the ``HTTP 302
Found'' redirection upon detection of the X-Accept-Datetime header,
only archival pages would be retrieved, and all links in those pages
would be interpreted as requests for Mementos. This effectively
happens for the servers under our control (the black flows labeled
[1], [2] and [4] in Figure \ref{fig:environment}). As described above,
other servers do not exhibit this behavior (the red flows [3] and [5]
in Figure \ref{fig:environment}). Implementing the remedial behavior
where the client itself takes care of the redirection turned out
not to be trivial in the Mozilla plug-in framework as it does not
support intercepting and modifying responses\footnote{\scriptsize{See
https://wiki.mozilla.org/Firefox/Projects/Network\_Error\_Pages}} (e.g.,
on 404 or 200 response codes).  The result is a time travel plug-in that
deals perfectly with URI-Rs of the servers under our control but not with
any others. We then decided to develop a time travel client that runs
on a server and is developed using the Apache mod\_python framework that
offered the required flexibility. The resulting gateway client handles all
flows of Figure \ref{fig:environment} correctly, and fully demonstrates
the potential of the DT-conneg framework. It is accessible via a web
form that allows entering URI-R and a datetime. Upon submitting the
time travel request, the gateway client (not the browser) fulfills the
DT-conneg requests, and once completely handled, returns the resulting
Memento page to the browser. In order to allow for continued time travel
of links in the page, they need to be rewritten to point at the gateway
client. This is merely an artifact of a server-side, not a browser-based,
implementation. This client also depicts the HTTP transactions that take
place during time travel, and allows inspecting the HTTP headers involved.

With the above components in place, an experimental environment results that effectively demonstrates the feasibility of web time travel using the Memento solution. Two clients, both admittedly with respective restrictions, allow navigating the past Web in very much the same way as the current Web is browsed; they seamlessly move across web servers and archives (CMS-style and web archives) using the HTTP protocol, extended with DT-conneg, to try and return a Memento that meets the client's preference. Due to the various by-proxy components, and the dynamic implementation of the aggregator, the navigation can often be slow. However, the navigations that involve the servers with full native support (flows [1] and [2] in Figure \ref{fig:environment}), those that bypass the aggregator (flows [1], [2] and [3]), and those for which the aggregator can respond from its limited cache (some flows [4] and [5]) perform noticeably faster, even using the gateway client. In addition, many well understood techniques including batch harvesting, caching, and recurrent refreshing are available to improve the performance of the aggregator and fundamentally improve response times.

\begin{figure*}
\begin{center}
\subfigure[http://lanlsource.lanl.gov/hello - flow 1 in Figure 4]{\label{lanl}\includegraphics[scale=0.16]{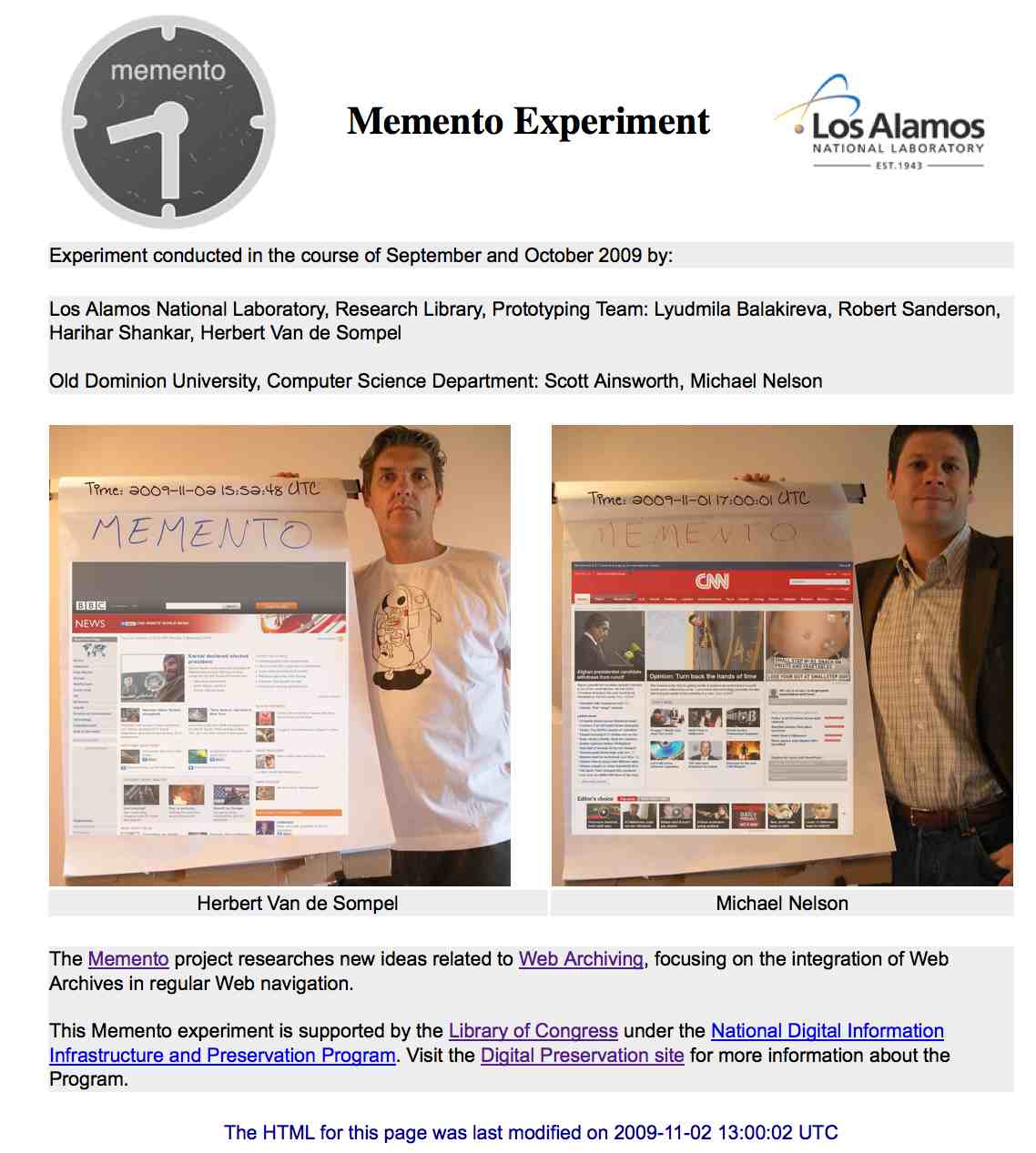}\includegraphics[scale=0.16]{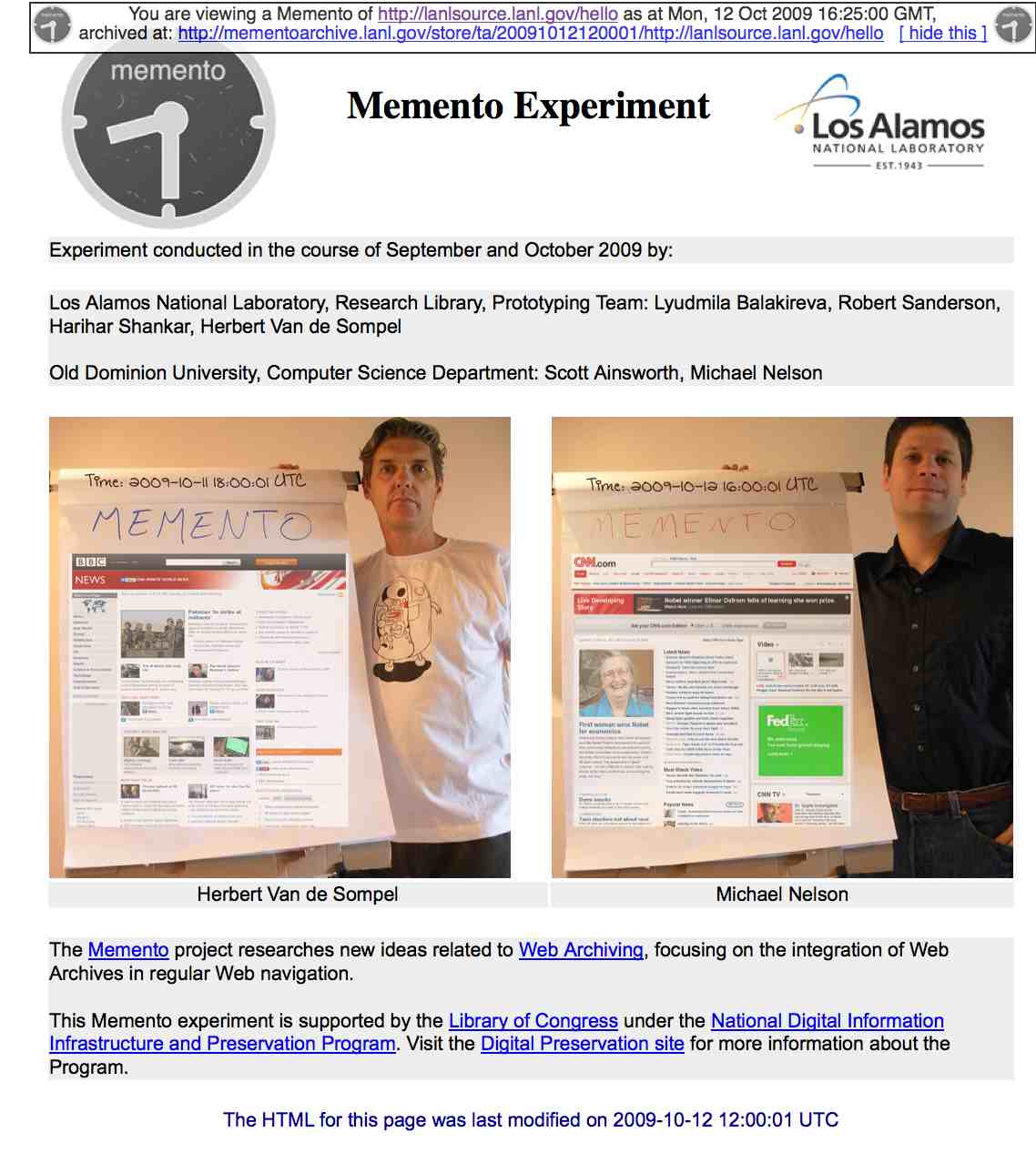}}
\subfigure[http://en.wikipedia.org/MS\_Oasis\_of\_the\_Seas - flow 3 in Figure 4]{\label{wikipedia}\includegraphics[scale=0.16]{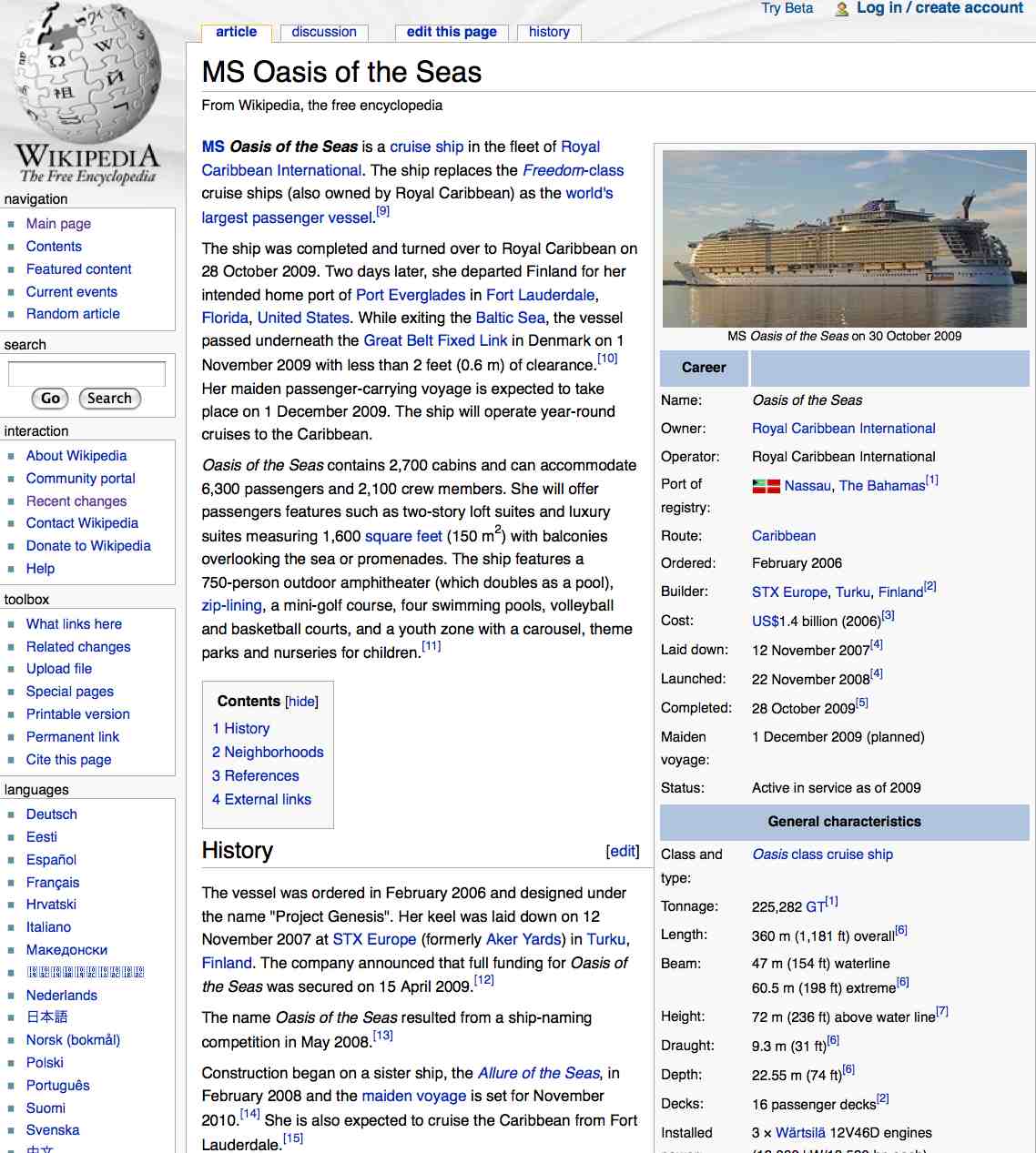}\includegraphics[scale=0.16]{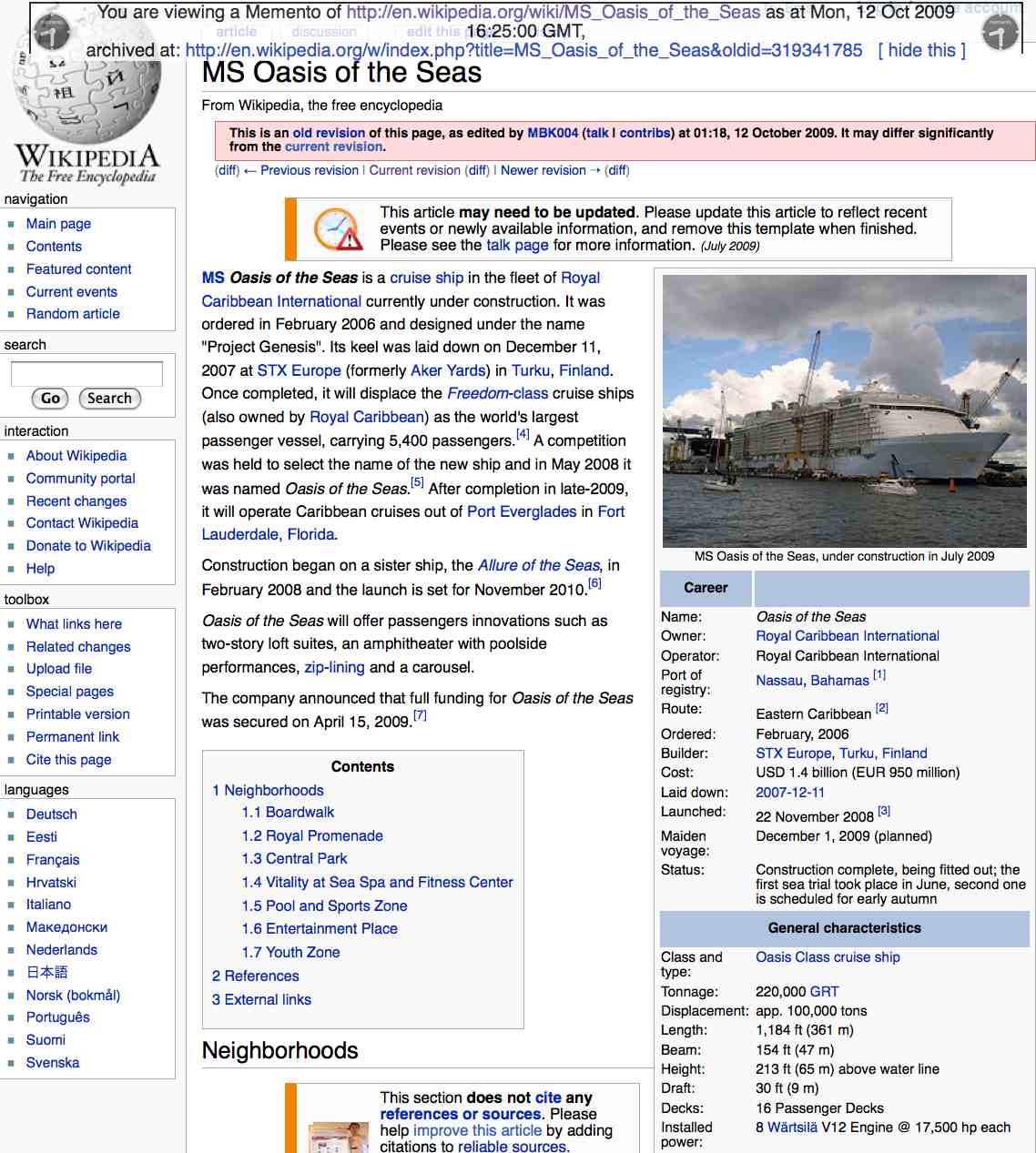}}
\subfigure[http://news.bbc.co.uk/ - flow 5 in Figure 4]{\label{bbc}\includegraphics[scale=0.16]{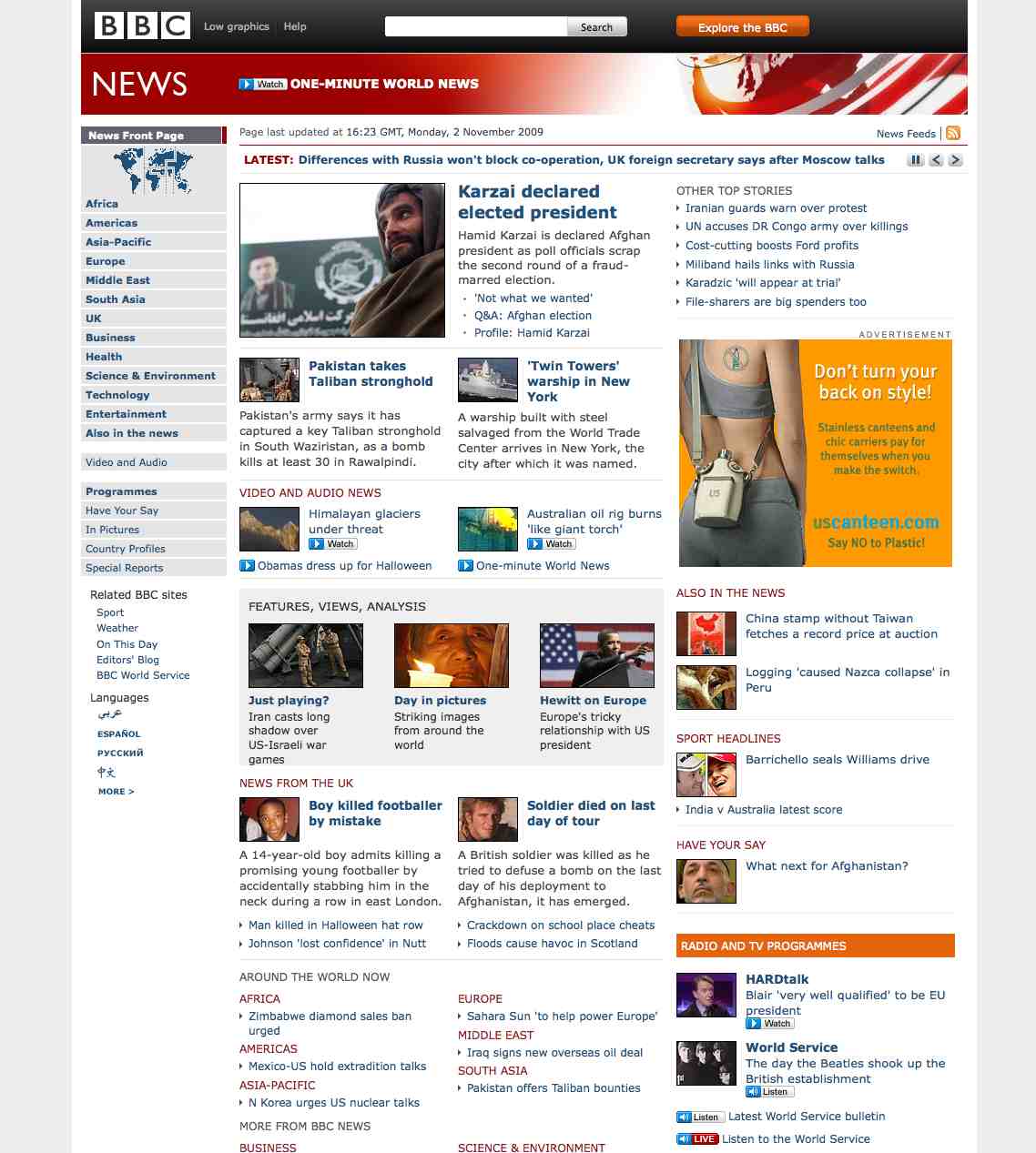}\includegraphics[scale=0.16]{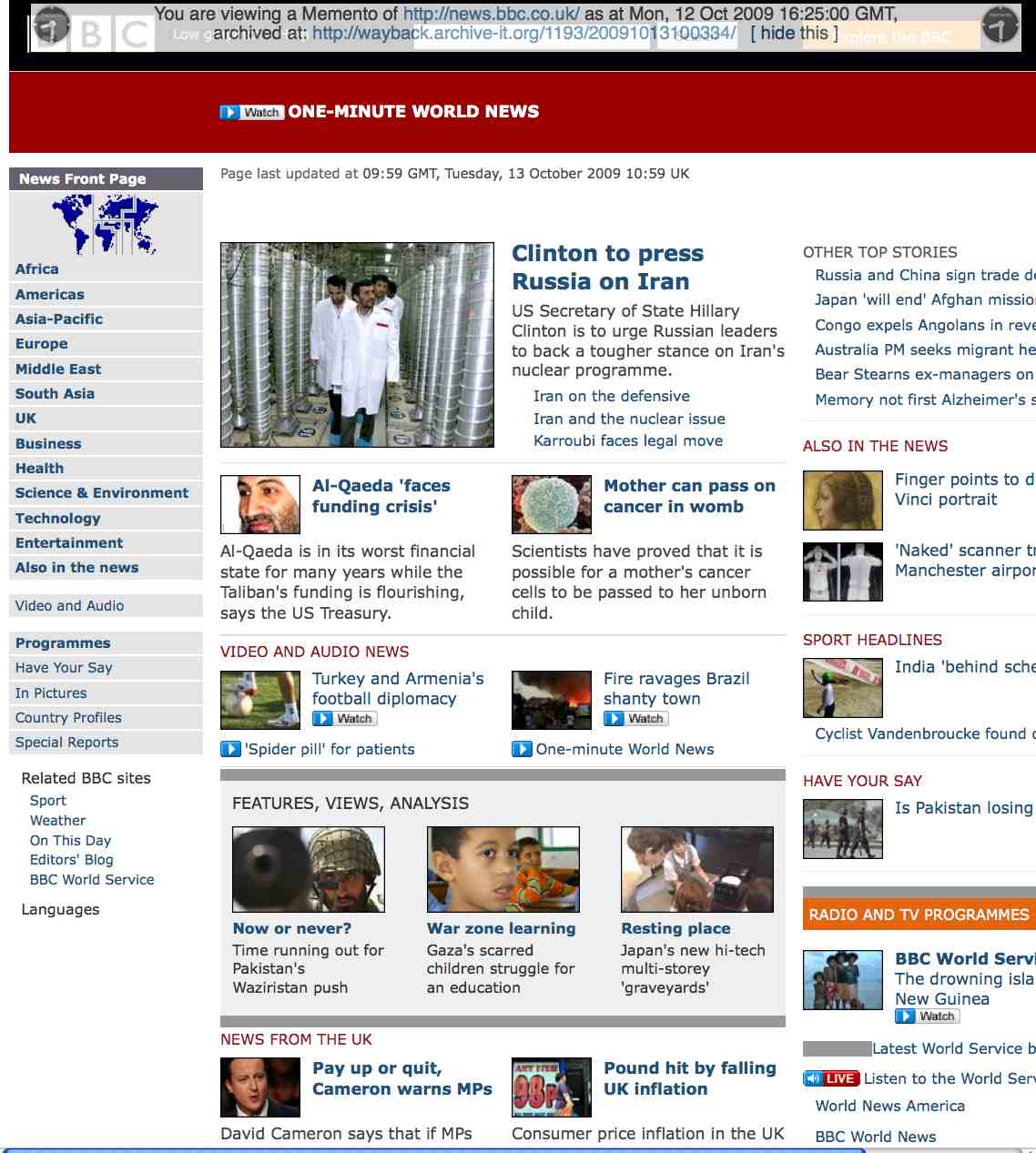}}
\caption{Browsing in real time (Mon, 02 Nov 2009 16:25:00 GMT) on the left and time travel (Mon,~12~Oct~2009~16:25:00~GMT) on the right.}
\label{fig:timetravel}
\end{center}
\end{figure*}

As an illustration of our results, Figure \ref{fig:timetravel} shows
two navigations conducted on November 2 2009, around 16:25:00 UTC:
one in real-time, and one in time travel mode with a datetime set to
October 12 2009 16:25:00 UTC.  The captions in the figure also indicate
the flow of the HTTP interactions in the experimental environment as
indicated in Figure \ref{fig:environment}.  The DT-conneg framework
allows a re-navigation of both in the future. It suffices to use these
respective datetimes as the DT-conneg value, and hope that archives
have records of the resources involved\footnote{\scriptsize{Try it at
http://mementoweb.org/demo/client1}}.

%

\section{Discussion}
\label{sec:issues}

In this section, we touch upon issues pertaining to the Memento solution that require further attention. The seamless integration of archives in regular web navigation provided by DT-conneg allows exploring novel ways to address common problems that result from the dynamics of the Web. Some of these have tentatively been explored in our experiment. Consider the following cases: 

\textit{A URI-R vanishes, but the server that used to serve it is still operational}: In this case, the server should still issue the redirect to a TimeGate upon detection of the DT-conneg request. This allows seamless access to a Memento of URI-R, even if the server no longer hosts the original. 

\textit{A domain vanishes}: The client is looking for a current representation of a URI-R that was hosted by the domain, but fails. The client resorts to interaction with other archives or with a TimeBundle aggregator and arrives at the most recent Memento of the resource. 

\textit{A domain is taken over by a new custodian}: The new custodian adheres to other policies regarding which archive to redirect a DT-conneg request. The client understands from the X-Archive-Interval returned by that archive of choice, that it does not cover the time range in which the previous custodian operated the domain. The client resorts to interaction with other archives or with a TimeBundle aggregator and arrives at an appropriate Memento.

Two aspects related to the integration of the proposed Memento solution into the existing Web infrastructure require attention. First, when issuing a request with an X-Accept-Datetime header to a server that hosts the original resource URI-R, all caches between the client and the server must be bypassed in order to avoid retrieving a current representation of URI-R. In our experiment, we enforced this behavior through a combination of two client request headers: ``Cache-Control: no-cache'' to force cache revalidation and ``If-Modified-Since: Thu, 01 Jan 1970 00:00:00 GMT'' to  make sure that revalidation fails.  Further research is required to find an alternative for this admittedly inelegant approach. Ideally, it should leverage existing caching practice but extend it in such a way that caches are only bypassed in DT-conneg when essential, but still used whenever possible (e.g., to deliver Mementos). Second, when it comes to listing variant resources in response headers, the DT-conneg framework cannot operate according to the letter of RFC 2295. Indeed, the RFC states: ``If a response from a transparently negotiable resource includes an Alternates header, this header MUST contain the complete variant list bound to the negotiable resource.'' This mandate is based on a perspective expressed in the RFC that ``it is expected that a typical transparently negotiable resource will have 2 to 10 variants, depending on its purpose.'' Clearly, TimeGates as proposed in the DT-conneg framework can have many more than 10 Mementos. We do not think this makes the conneg framework inapplicable to the datetime dimension, but rather we believe DT-conneg introduces a challenge that the authors of the RFC did not anticipate ten years ago. As described, we propose a solution based on a sizeable Alternates header combined with an HTTP Link header that leads to a complete list of variants; other options should be explored.

An interesting characteristic of the DT-conneg framework requires more explicit attention. When requesting a Memento for a page that contains links to external pages, or embedded resources such as images or videos, each of those are requested with DT-conneg from the respective servers that host/hosted the URI-R of those resources. This is a core characteristic that the proposed time travel framework shares with regular web navigation. It should be noted that this is not the current behavior of pages stored in web archives. Indeed, in order to avoid filling out an archived page with current representations of embedded resources, web archives rewrite URIs in archived pages to point back into the archive at archived representations of those resources. The same happens with links in archived pages, effectively turning the archive into an island isolated from the rest of the Web. The upside of this approach is that archived pages are self-contained: the page and its embedded resources were typically crawled around the same time and hence the archived page is likely to be a faithful reconstruction of what the original looked like at the time of the crawl. The drawback of the approach is that navigation is restricted to the archive's island. Navigating beyond it to obtain an archived version of a linked resource that is not available in the archive but might be available elsewhere on the Web, is not possible. Further exploration is required to arrive at a strategy for web archives that would at the same time adhere to the self-containedness principle and allow external navigation using the DT-conneg framework when beneficial. 

Another challenge pertains to selecting a Memento that best meets the client's conneg preferences. There are two aspects to this problem. The first relates to the archival datetime of the Memento that an archive should return in response to a datetime expressed by a time travel client. For certain archives the choice is straightforward. Indeed, transactional archives and servers such as Wikipedia know exactly during which time interval a certain Memento functioned as the active representation of URI-R (cf. the X-Datetime-Validity discussed in Section \ref{sec:headers}). Hence, they can return the Memento that was active at the datetime specified by the client. However for resources not hosted by such servers, it will be rare that any archive has a Memento that perfectly matches the client's preference. In this case, an archive (or a TimeBundle aggregator) must make a choice. A typical approach used by existing web archives is to choose the Memento that is the ``closest'' in time, regardless of whether its archival datetime is before or after the requested datetime. But this approach is challenged when pages have embedded resources.  The more resources required to render a page, the more variation there will be between the requested datetime and the archival datetimes of available Mementos. As a matter of fact, when not being sensible about the selection of Mementos, the resulting page may never actually have existed. A second challenge relates to multi dimensional conneg that involves the datetime dimension. Current conneg algorithms\footnote{\scriptsize{See, http://httpd.apache.org/docs/2.2/content-negotiation.html}} deal with variant selection in the dimensions specified in RFC 2295. These would need to be revised to include the datetime dimension: if a client requests an HTML Memento for a specific datetime, but only a pdf is available, what should the archival server do?  Research is required to explore both problems.  

\section{Related Work}
\label{sec:related}

The goal of adding a temporal aspect to web navigation has been explored in projects that focus on user interface enhancement. The Zoetrope project \cite{1449756} provides a rich interface for
querying and interacting with a set of archived versions of selected seed pages. The interface leverages a local archive that is assembled by frequently polling those seed pages. The Past Web Browser
\cite{1149969} provides a simpler level of interaction
with changing pages, but it is restricted to navigating existing web archives such as
the Internet Archive. And DiffIE is a plug-in for Internet Explorer that emphasizes
web content that changed since a user's previous visit by leveraging a dedicated client cache \cite{1622221}. None of these projects propose protocol enhancements but rather use ad-hoc techniques to achieve their goals. All could benefit from DT-conneg as a standard mechanism for accessing prior representations of resources.

Some projects have dealt with the problem of disappeared web pages and finding archived or replacement copies on the Web.  The use of lexical signatures as search engine query terms was proposed as a way to find content that had moved from its original URI \cite{1028101, phelps:robust}. This approach was later applied to search for content in web archives \cite{opal:ht06, lexsig:ecdl08}. Also, when a ``HTTP 404 Not Found'' occurs, the ErrorZilla FireFox plug-in\footnote{\scriptsize{https://addons.mozilla.org/en-US/firefox/addon/3336}} presents a user with a search page allowing her to find disappeared pages in web archives, and the UK National Archive's server plug-in redirects the client to an archive of its choice. As suggested (Section \ref{sec:issues}), in the DT-conneg framework a client could intelligently react to 404s, and when doing so leverage available re-finding approaches. 

To the best of our knowledge, very little research has explored a protocol-based solution to augment the Web with time travel capabilities. TTApache \cite{dyreson:managing} introduced a modified version of
Apache that stored archived representations in a local transactional archive
(similar to the configuration illustrated in Figure \ref{fig:scenario1}).
Ad-hoc RPC-style mechanisms were used to access archived representations given the URI of their original, e.g. ``page.html?02-Nov-2009'' and \\``page.html?now''. This approach reveals the local scope of the problem addressed by TTApache, as opposed to the global perspective taken by the proposed DT-conneg framework. Indeed, the query components are issued against a specific server, and are not maintained when a client moves to another server as is the case with the X-Accept-Datetime header of DT-conneg. TTApache also allowed addressing archived representations using version numbers in query \\components rather than datetimes. This capability is similar to the deprecated ``Content-Version'' header field from RFC 2068 \cite{rfc2068} and other, similar expired proposals (e.g., \cite{meta-level-draft}). Such versioning features have not found wide-spread adoption, presumably because their address space is tied to a specific resource or server, and not universal like the datetime of DT-conneg.

\section{Conclusions}
\label{sec:conclusions}

In Web Archiving \cite{masanes:web-archiving-book}, Julien Masan{\`e}s expresses a vision of a global grid of web archives realized by interconnecting existing and future ones:

\begin{quote}
Such a grid should link Web archives so that they together form one global navigation space like the live Web itself. This is only possible if they are structured in a way close enough to the original Web and if they are openly accessible.
\end{quote}
   
We could not agree more, and feel that our Memento solution presents a significant step towards achieving this vision. But our approach reaches beyond it. Indeed, the navigation space that results from our proposal is not ``like the live Web itself'', it \textbf{is} the Web itself, as regular navigation and time travel are integrated. Also, it does not restrict the global archival grid to web archives but incorporates servers (such as CMS) on the live Web that host archival content. The Memento solution is  capable of realizing this, and does not disrupt firmly established HTTP practice. Rather, it adds to it an orthogonal time dimension. Moreover, the Memento solution does not disrupt existing web archives or their established operating principles, but leverages both by tightly integrating them into the web. Time travel can be ours.
 
\section{Acknowledgments}
This work sponsored in part by the Library of Congress.

%
\bibliographystyle{abbrv}
\bibliography{memento_20091102.bib}  

\begin{thebibliography}{10}

\bibitem{1449756}
E.~Adar, M.~Dontcheva, J.~Fogarty, and D.~S. Weld.
\newblock Zoetrope: interacting with the ephemeral web.
\newblock In {\em UIST '08: Proceedings of the 21st annual ACM symposium on
  User interface software and technology}, pages 239--248, 2008.

\bibitem{1518909}
E.~Adar, J.~Teevan, and S.~T. Dumais.
\newblock Resonance on the web: web dynamics and revisitation patterns.
\newblock In {\em CHI '09: Proceedings of the 27th international conference on
  Human factors in computing systems}, pages 1381--1390, 2009.

\bibitem{1498837}
E.~Adar, J.~Teevan, S.~T. Dumais, and J.~L. Elsas.
\newblock The web changes everything: understanding the dynamics of web
  content.
\newblock In {\em WSDM '09: Proceedings of the Second ACM International
  Conference on Web Search and Data Mining}, pages 282--291, 2009.

\bibitem{1555455}
A.~Anand, S.~Bedathur, K.~Berberich, R.~Schenkel, and C.~Tryfonopoulos.
\newblock Everlast: a distributed architecture for preserving the web.
\newblock In {\em JCDL '09: Proceedings of the 9th ACM/IEEE-CS joint conference
  on Digital libraries}, pages 331--340, 2009.

\bibitem{linkeddata}
C.~Bizer, R.~Cyganiak, and T.~Heath.
\newblock How to publish linked data on the web, 2007.
\newblock http://sites.wiwiss.fu-berlin.de/bizer/pub/LinkedDataTutorial/.

\bibitem{brown:archiving-websites}
A.~Brown.
\newblock {\em Archiving Websites: A practical guide for information management
  professionals}.
\newblock Facet Publishing, 2006.

\bibitem{cooper:infomonitor}
B.~F. Cooper and H.~Garcia-Molina.
\newblock Infomonitor: Unobtrusively archiving a {World Wide Web} server.
\newblock {\em International Journal on Digital Libraries}, 5(2):106--119,
  April 2005.

\bibitem{dyreson:managing}
C.~E. Dyreson, H.~Lin, and Y.~Wang.
\newblock Managing versions of web documents in a transaction-time web server.
\newblock In {\em WWW '04: Proceedings of the 13th international conference on
  World Wide Web}, pages 422--432, 2004.

\bibitem{fetterly:large-scale}
D.~Fetterly, M.~Manasse, M.~Najork, and J.~Wiener.
\newblock A large-scale study of the evolution of web pages.
\newblock In {\em WWW '03: Proceedings of the 12th international conference on
  World Wide Web}, pages 669--678, 2003.

\bibitem{rfc2068}
R.~Fielding, J.~Gettys, J.~Mogul, H.~Frystyk, and T.~Berners-Lee.
\newblock Hypertex transfer protocol -- {HTTP}/1.1, {Internet RFC-2068}, 1997.

\bibitem{fitch2003web}
K.~Fitch.
\newblock {Web site archiving: an approach to recording every materially
  different response produced by a Website}.
\newblock In {\em 9th Australasian World Wide Web Conference, Sanctuary Cove,
  Queensland, Australia, July}, pages 5--9, 2003.

\bibitem{sitemaps}
{Google}, {Microsoft}, and {Yahoo}.
\newblock Sitemaps {XML} format, 2008.
\newblock http://www.sitemaps.org/protocol.php.

\bibitem{opal:ht06}
T.~L. Harrison and M.~L. Nelson.
\newblock Just-in-time recovery of missing web pages.
\newblock In {\em HYPERTEXT '06: Proceedings of the Seventeenth ACM Conference
  on Hypertext and Hypermedia}, pages 145--156, 2006.

\bibitem{rfc2295}
K.~Holtman and A.~Mutz.
\newblock Transparent content negotiation in {HTTP, Internet RFC-2295}, 1998.

\bibitem{archWWW}
I.~Jacobs and N.~Walsh.
\newblock Architecture of the world wide web, volume one.
\newblock Technical Report W3C Recommendation 15 December 2004, W3C, 2004.

\bibitem{1149969}
A.~Jatowt, Y.~Kawai, S.~Nakamura, Y.~Kidawara, and K.~Tanaka.
\newblock Journey to the past: proposal of a framework for past web browser.
\newblock In {\em HYPERTEXT '06: Proceedings of the seventeenth conference on
  Hypertext and hypermedia}, pages 135--144, 2006.

\bibitem{lexsig:ecdl08}
M.~Klein and M.~L. Nelson.
\newblock Revisiting lexical signatures to (re-)discover web pages.
\newblock In {\em ECDL '08: Proceedings of the 12th European Conference on
  Research and Advanced Technology for Digital Libraries}, pages 371 -- 382,
  2008.

\bibitem{koehler:web-page}
W.~Koehler.
\newblock Web page change and persistence --- a four-year longitudinal study.
\newblock {\em Journal of the American Society for Information Science and
  Technology}, 53(2):162--171, 2002.

\bibitem{379449}
C.~Lagoze and H.~{Van de Sompel}.
\newblock The {Open Archives Initiative}: building a low-barrier
  interoperability framework.
\newblock In {\em JCDL '01: Proceedings of the 1st ACM/IEEE-CS Joint Conference
  on Digital Libraries}, pages 54--62, 2001.

\bibitem{1557077}
J.~Leskovec, L.~Backstrom, and J.~Kleinberg.
\newblock Meme-tracking and the dynamics of the news cycle.
\newblock In {\em KDD '09: Proceedings of the 15th ACM SIGKDD international
  conference on Knowledge discovery and data mining}, pages 497--506, 2009.

\bibitem{w3c:HTTP14}
R.~Lewis.
\newblock Dereferencing {HTTP URIs}.
\newblock Technical Report Draft Tag Finding 04 October 2007, 2007.

\bibitem{masanes:web-archiving-book}
J.~Masan{\`e}s.
\newblock {\em Web Archiving}.
\newblock Springer-Verlag, 2006.

\bibitem{HTTPLink}
M.~Nottingham.
\newblock Web linking, {Internet Draft draft-nottinghamgm-http-link-header-06},
  2009.

\bibitem{rfc:4287}
M.~Nottingham and R.~Sayre.
\newblock The {Atom} syndication format, {Internet RFC-4287}, 2005.

\bibitem{988674}
A.~Ntoulas, J.~Cho, and C.~Olston.
\newblock What's new on the web?: the evolution of the web from a search engine
  perspective.
\newblock In {\em WWW '04: Proceedings of the 13th international Conference on
  World Wide Web}, pages 1--12, 2004.

\bibitem{Obendorf:chi07}
H.~W. E.~H. Obendorf, Hartmut and M.~Mayer.
\newblock Web page revisitation revisited: Implications of a long-term
  click-stream study of browser usage.
\newblock In {\em CHI '07: Proceedings of the 25th international conference on
  Human factors in computing systems}, pages 597--606, 2007.

\bibitem{meta-level-draft}
K.~Ota, K.~Takahashi, and K.~Sekiya.
\newblock Version management with meta-level links via {HTTP}/1.1, {Internet
  Draft draft-ntt-http-version-00}, 1996.

\bibitem{1028101}
S.-T. Park, D.~M. Pennock, C.~L. Giles, and R.~Krovetz.
\newblock Analysis of lexical signatures for improving information persistence
  on the {World Wide Web}.
\newblock {\em ACM Transactions on Information Systems}, 22(4):540--572, 2004.

\bibitem{phelps:robust}
T.~A. Phelps and R.~Wilensky.
\newblock Robust hyperlinks cost just five words each.
\newblock Technical Report UCB/CSD-00-1091, EECS Department, University of
  California, Berkeley, 2000.

\bibitem{chungwwa:webarchiving}
H.~C. Rao, Y.~Chen, and M.~Chen.
\newblock A proxy-based personal web archiving service.
\newblock {\em SIGOPS Operating Systems Review}, 35(1):61--72, 2001.

\bibitem{tauscher1997}
L.~Tauscher and S.~Greenberg.
\newblock How people revisit web pages: Empirical findings and implications for
  the design of history systems.
\newblock {\em International Journal of Human-Computer Studies}, 47(1), 1997.

\bibitem{1622221}
J.~Teevan, S.~T. Dumais, D.~J. Liebling, and R.~L. Hughes.
\newblock Changing how people view changes on the web.
\newblock In {\em UIST '09: Proceedings of the 22nd annual ACM symposium on
  User interface software and technology}, pages 237--246, 2009.

\bibitem{w3c:fragment}
R.~Troncy, J.~Jansen, Y.~Lafon, E.~Mannens, S.~Pfeiffer, and D.~V. Deursen.
\newblock Use cases and requirements for media fragments, {W3C Working Draft 30
  April 2009}, 2009.

\bibitem{ldow2009:ore}
H.~Van~de Sompel, C.~Lagoze, M.~L. Nelson, S.~Warner, R.~Sanderson, and
  P.~Johnston.
\newblock Adding escience assets to the data web.
\newblock In {\em Proceedings of the Linked Data on the Web Workshop (LDOW
  2009)}, 2009.

\end{thebibliography}
\balancecolumns 
\end{document}